\documentclass[a4paper,twocolumn]{article}
\usepackage{hyperref}
\usepackage{graphicx}
\usepackage{amsmath}
\usepackage{amsthm}
\usepackage{booktabs}

\usepackage[utf8]{inputenc}
\usepackage{tikz}
\usepackage{calc}
\usepackage{todonotes}
\usepackage{xspace}

\theoremstyle{definition}
\newtheorem{definition}{Definition}

\usepackage{comment}
\usepackage{bold-extra}
\usepackage{textcomp}
\usepackage{pgfplots}
\pgfplotsset{width=10cm,compat=1.9}

\theoremstyle{plain}

\newcommand{\NP}{\textsf{NP}\xspace}

\DeclareMathOperator{\KTdist}{KT-distance}
\DeclareMathOperator{\row}{row}
\DeclareMathOperator{\col}{col}
\DeclareMathOperator{\rank}{rank}

\newcommand{\abs}[1]{\vert #1 \vert}

\newcommand{\SteepestDescentSolverName}{\textsc{SteepDes}\xspace}
\newcommand{\SimulatedAnnealingSamplerName}{\textsc{SimAnl}\xspace}
\newcommand{\LeapHybridSamplerName}{\textsc{Hybrid}\xspace}
\newcommand{\DWaveSamplerName}{\textsc{QAnl}\xspace}
\newcommand{\DWaveName}{{D-Wave}\xspace}

\newcommand{\LocalSearch}{\textsc{LocSearch}\xspace}
\newcommand{\Borda}{\textsc{Borda}\xspace}
\newcommand{\QuickSort}{\textsc{QuickSort}\xspace}

\usepackage{authblk}
\title{Heuristic for Diverse Kemeny Rank Aggregation based on Quantum Annealing}
\author[1]{Sven Fiergolla}
\author[1]{Kevin Goergen}
\author[1]{Patrick Neises}
\author[2]{Petra Wolf}
\affil[1]{University of Trier, Germany, \texttt{\{s4svfier, s4kegoer,s4paneis\}@uni-trier.de}}
\affil[2]{University of Bergen, Norway, \texttt{mail@wolfp.net}}
\date{}  
\begin{document}
	
	\maketitle
\begin{abstract}
	The \textsc{Kemeny Rank Aggregation} (\textsc{KRA}) problem is a well-studied problem in the field of Social Choice with a variety of applications in many different areas like databases and search engines. 
	Intuitively, given a set of votes over a set of candidates, the problem asks to find an aggregated ranking of candidates that minimizes the overall dissatisfaction concerning the votes.
	Recently, a diverse version of \textsc{KRA} was considered which asks for a sufficiently diverse set of sufficiently good solutions. 
	The framework of \emph{diversity of solutions} is a young and thriving topic in the field of artificial intelligence. The main idea is to provide the user with not just one, but with a set of different solutions, enabling her to pick a sufficiently good solution that satisfies additional subjective criteria that are hard or impossible to model. 
%	Thereby, the set of solutions ideally represents the solution space. 

	In this work, we use a \emph{quantum annealer} to solve the \textsc{KRA} problem and to compute a representative set of solutions. Quantum annealing is a meta search heuristic that does not only show promising runtime behavior on currently existing prototypes but also samples the solutions space in an inherently different way, making use of quantum effects.
	We describe how \textsc{KRA} instances can be solved by a quantum annealer and provide an implementation as well as experimental evaluations.
	As existing quantum annealers are still restricted in their number of qubits, we further implement two different data reduction rules that can split an instance into a set of smaller instances.
%	 which can be solved independently.
	In our evaluation, we compare classical heuristics that allow to sample multiple solutions such as simulated annealing and local search with quantum annealing performed on a physical quantum annealer. We compare runtime, quality of solution, and diversity of solutions, with and without applying preceding data reduction rules. While our aim is to compute a \emph{diverse set} of solutions, we also include classical heuristics that only compute a single solution in our experiments.
	
\end{abstract}

\section{Introduction}

%\todo[inline]{Using different heuristics that allow to sample multiple solutions to get a heuristic for the diverse KRA Problem. One of them is Quantum Annealing, then intro of Quantum.}
Estimating the potential of quantum computers gains more and more importance as their technology develops rapidly. There are two different approaches for utilizing quantum effects to solve computational problems: the \emph{Quantum Circuit Model} (an approach followed by Google and NASA~\cite{arute2019quantum}, and IBM~\cite{steffen2011quantum}) and the \emph{Adiabatic Quantum Model} (an approach dominated by D-Wave Systems~\cite{king2022coherent}).
While both models are computationally equivalent in the sense that they can simulate each other with only a polynomial overhead~\cite{farhi2000quantum,DBLP:journals/siamrev/AharonovDKLLR08} their physical realizations are quite different.
The quantum circuit model aims at building a universal quantum computer, it requires the design of novel quantum algorithms to make use of quantum effects. 
As quantum algorithms are quite unintuitive, only few quantum algorithms are known that provide significant benefits over classical algorithms (for instance, Grover's algorithm searching in unstructured data with a quadratic speedup~\cite{DBLP:conf/stoc/Grover96}, and Shor's algorithm factoring a number in polynomial time~\cite{DBLP:conf/focs/Shor94}).
In contrast, the Adiabatic Quantum Model has been realized by D-Wave as a solution method called \emph{quantum annealing} (QA) to solve \NP-hard optimization problems. The quantum annealers from D-Wave are designed to solve a native problem called the \emph{Ising Model}\footnote{The Ising Model problem is \NP-complete~\cite{DBLP:conf/stoc/Istrail00,cipra2000ising} and remains \NP-complete restricted to the class of instances realizable on a D-Wave quantum annealer~\cite{bunyk2014architectural}.} on their hardware chip.
%The Ising Model aims to find, in a system of particles, a spin orientation of the particles in the lowest energy configuration possible. 
%The problem
The Ising Model is very closely related to the \emph{quadratic unconstrained binary optimization} (QUBO) problem which is more accessible from the perspective of a computer scientist and, hence, will be considered in this paper.
The QUBO problem provides us with a nice interface that allows to formulate problems as a classical optimization problem while still utilizing quantum effects to speed up the computation. 
We discuss the QUBO problem in detail later.

Intuitively, in the Adiabatic Quantum Model, the system is transformed from an initial low energy state, which is easy to prepare, into another low energy state 
%of a target Hamiltonian 
such that the final state corresponds to minimizing a target function. If this transition is performed slowly enough and at very low temperatures ($\leq$20mK), the system stays in the lowest energy state during the transformation and we can measure an optimal solution from the final state.
The Adiabatic Quantum Model was initially proposed in~\cite{farhi2001quantum}
%, see also~\cite{farhi2000quantum}, 
and soon found its application in neural networks~\cite{DBLP:conf/icann/KinjoSN03}. 

Quantum annealing is a meta search heuristic that can be implemented in the native instruction set of the Adiabatic Quantum Model.
The most developed realization in form of a quantum annealer is the D-Wave Advantage system with 5,000+ qubits~\cite{willsch2022benchmarking}. The quantum computer from D-Wave also allows for Quantum Machine Learning applications \cite{hu2019quantum,adachi2015application}. The company rapidly develops new hardware chips for quantum annealers, which does not come without some difficulties concerning API stability.
% and comparability of results. 
For more details on the technical realization see~\cite{bunyk2014architectural,harris2010experimental,johnson2011quantum} and for a general introduction to the Adiabatic Quantum Model see~\cite{DBLP:series/synthesis/2014McGeoch}.

The Adiabatic Quantum Model differs from classical simulations of thermal annealing~\cite{kirkpatrick1983optimization} in the sense that the computation starts with a superposition of all states with minimal energy, thereby exploiting \emph{quantum parallelism}, and can further \emph{tunnel} through high energy areas in the solution space while classical approaches need to climb over those high energy regions. 
Simulations have shown that thereby the Adiabatic Quantum Model has the potential of outperforming classical simulated annealing~\cite{morita2008mathematical,ohzeki2011quantum,crosson2016simulated}. Further, experimental results show that quantum annealers can find local valleys in the solution space that are not found by simulated annealers~\cite{koshka2020comparison}.
%Further, concerning quantum annealing, in the process of modifying the system from its initial to its target state, entanglements between quantum states can occur, as observed in~\cite{lanting2014entanglement} leading to an observed bimodality in the probability of solving an instance optimally, that differs from the classical case~\cite{boixo2013experimental,boixo2014evidence}. 
This indicates that instances that cannot be solved efficiently with classical simulated annealing might be efficiently solvable on a quantum annealer.

Due to its stochastic nature, the quantum annealing process must be executed and measured multiple times to get good solutions. But while this looks like a weakness of the approach, it actually comes with a benefit:
While the system cools down, one can imagine it as being in a superposition of all local minima. Only once measured, the system collapse to one minima. By repeating the process we can sample the different minima of the solution space and do not get stuck in one local minima as it is often the case for gradient descent approaches. 
The multiple solutions automatically sampled further allow us to get a representation of the solution space, an aim that is classically modeled by the notion of \emph{diversity of solutions}, a recent upcoming trend in artificial intelligence~\cite{PetitTrapp2019,BasteEtAl19,BasteFJMOPR20,ingmar2020modelling,fomin2020diverse}.

While, traditionally, one is interested in getting \emph{some} optimal solution to a computational problem, this might not be sufficient in practice when side constraints are present, that are hard to model. 
Examples for those subjective criteria are aesthetic, economic, political, or environmental criteria.
Due to its subjective vague nature, it can be impossible to model those criteria on the level of the problem formulation.
Hence, it can be preferable for the user to choose a slightly less optimal solution that suits her subjective criteria better. 
In order to provide the user with the ability to choose the solution that fits her needs best, it is preferable to provide a reasonably sized solution set of sufficiently diverse solutions.
But formalizing the diversity of a solution set is a non-trivial task, see the discussion in~\cite{BasteFJMOPR20,baste2019fpt}.
Intuitively, diversity serves as a measure of how representative a set of solutions is among the solution space. 
%While classically, several different measures have been used to formalize diversity, see \cite[Sec.~3]{DBLP:conf/ppsn/UlrichBT10}, 
A quantum annealer naturally provides us with a set of solutions as it samples the solution space.
% that authentically represent the solution space. 
Then, classical post-processing can be applied to select a subset of these solutions suitable for the criteria at hand.

Considering diverse alternatives is of relevance while selecting several good committees such that each committee member is not overloaded with work, as described in~\cite{BreKacNie2020}.
Electing committees, or finding a ranking that agrees with a set of votes as much as possible, are key problems in the field of Social Choice. 
One of the most studied problems in the field is the \textsc{Kemeny Rank Aggregation} (\textsc{KRA}) problem.
Here, a set of votes over a set of candidates is given and the task is to find a ranking of the candidates that maximizes the satisfaction of the voters. 
The problem is of relevance in a variety of different areas such as similarity search
and classification in high dimensional databases~\cite{DBLP:conf/sigmod/FaginKS03}, constructing genetic maps in bioinformatics~\cite{DBLP:journals/tcbb/JacksonSA08}, aggregating search results and spam detection in a search engine~\cite{DBLP:conf/www/DworkKNS01}, etc.
The popularity of \textsc{KRA} is due to the unique property of the Kemeny rule of being neutral, consistent, and satisfying the Condorcet property~\cite{young1978consistent}.
%Since its introduction in the late fifties, \cite{Kemeny59}, the framework of preference list aggregation is still of great interest nowadays as underlined by articles describing it for the interested public audience; see \cite{FarTim2019}.
%The framework of preference list aggregation was introduced by 
%Kemeny in the late fifties \cite{Kemeny59} and is still of great interest nowadays as underlined by articles describing preference aggregation issues for the interested public audience; see \cite{FarTim2019}.

The \textsc{KRA} problem is \NP-complete~\cite{bartholdi1989voting} and remains \NP-hard when restricted to only four votes~\cite{DBLP:conf/www/DworkKNS01,DBLP:journals/dm/BiedlBD09}. This hardness motivates the studies of heuristics and approximation algorithms for \textsc{KRA}, see~\cite{DBLP:conf/www/DworkKNS01,DBLP:journals/jacm/AilonCN08,DBLP:journals/mor/ZuylenW09}, as well as parameterized algorithms~\cite{DBLP:journals/tcs/BetzlerFGNR09,DBLP:conf/iwpec/Simjour09,DBLP:conf/fsttcs/ArrighiFO020,DBLP:journals/aamas/BetzlerBN14}.
The practical relevance in the field of artificial intelligence is also highlighted by a series of experimental studies of heuristics and approximations of the \textsc{KRA} problem~\cite{DBLP:journals/mss/AliM12,DBLP:conf/aaai/DavenportK04} as well as experimental studies of lower bounds for exact solutions of \textsc{KRA}~\cite{DBLP:conf/aaai/ConitzerDK06,DBLP:conf/alenex/SchalekampZ09}.
%\todo{say in experiment section that data and solver are not available}
A parameterized algorithm for a diverse version of \textsc{Kemeny Rank Aggregation} recently appeared at IJCAI~\cite{DBLP:conf/ijcai/ArrighiFLO021}.

In this work, we utilize the Adiabatic Quantum Model to get heuristics for the \textsc{Diverse Kemeny Rank Aggregation} problem that provide us with a set of samples of the solution space.
%good solutions that naturally\todo{todo} represent the solutions space. 
For this, we model \textsc{KRA} as a quadratic unconstrained binary optimization problem.
%, an equivalent formulation of the native Ising Model problem solved by quantum annealers. 
Then, we compare the quantum based heuristics with classical heuristics that allow to sample different solutions such as simulated annealing and local search, as well as classical heuristics that compute deterministically only a single solution.

As quantum computer
%, as well as quantum computer based on the gate model, 
are still quite restricted in the number of available qubits (currently 5,000+ qubits), large instances are still a challenge to any quantum computer. To counter this fact, we implement several data reduction rules as a  preprocessing step. Those rules allow us to break an instance into smaller instances that can be solved independently. A final solution is then aggregated from the solutions of the smaller instances.
% We describe the data reduction rules in more detail later.

\begin{comment}
Quadratic Unconstrained Binary Optimization (QUBO) is an NP-hard combinatorial optimization problem. Since it is a very natural formulation for many problems encountered in computer science, economics and other fields, efforts have been made to find efficient ways to solve instances of QUBO. An especially promising way to achieve this is utilizing Quantum Annealers, the physical architecture of which matches the structure of QUBO nicely.

Quantum Annealing is one of multiple possible internal structures for quantum computers. It utilizes quantum fluctuations, in order to identify the globally lowest energy state of a system. This technique makes a natural fit for optimization problems, such as QUBO, in which it is often desirable to find global minima.

Due to this beneficial way of solving instances of QUBO, polynomial-time reductions from many different problems to QUBO and QUBO formulations for natural problems have been found over the last few years.

In this paper, we present such a formulation for Kemeny Rank Aggregation (KRA). KRA is an electoral system, which makes use of ranked voting in order to aggregate the most popular ranking of candidates in an election. Beyond use in actual elections, KRA can also be used to aggregate rankings in other contexts like sorting search results by relevance. In this paper we focus on an electoral rule set in which each voter has to rank all candidates and different candidates cannot be assigned the same rank.
\end{comment}

This work is structured as follows. First, we give formal definitions for the QUBO and \textsc{KRA} problem. Then, we describe the data reduction rules we used as well as the QUBO formulation obtained for \textsc{KRA}. We continue with the experimental part of the paper and introduce the considered solvers.
% where the first three are provided by D-Wave: steepest decent, simulated annealing, quantum annealing, local search, Borda, and a solver based on quick-sort.
% three different types of solvers provided by D-Wave: classical mathematical solvers, simulated annealing, and quantum annealing. 
In our experiments, we compare the runtime and obtained solution quality for all the solvers, and diversity of solutions for solvers that support computing multiple solutions. Thereby, we compare three different preprocessing modes: no preprocessing, applying the $\leq_{3/4}$-majority rule exhaustively, and applying the extended Condorcet rule. 
%As the available real-world data-sets where already solved by the implemented data reduction rules, see the discussion in~\cite{DBLP:journals/aamas/BetzlerBN14}, we focused our experiments on randomly generated instances. 
%We observe that the $\leq_{3/4}$-majority rule does not reduce any instance in our data-set. In contrast, the Condorcet rule reduces nearly every instance and further increases the solution quality. 
%Details on our observation can be found at the end of this paper.
We present our observations at the end of this paper.
Our implementation and the used data-sets are available as supplementary material.

%The main benefit of our results is a proof of concept that we can solve \textsc{KRA} using a Quantum Annealer and that we can outperform classical solvers on the QUBO Model of \textsc{KRA}.

\section{Preliminaries}
Let $Q$ be an $n\times n$ matrix. We write $Q_{ij}$ for the entry in row~$i$ and column~$j$ of $Q$. 
Let $n$ be a non-negative integer. We denote $[n] = \{1, 2, \dots n\}$. Consequently, $[0] = \emptyset$.
\subsection{Quadratic Unconstrained Binary Optimization}
We now give a formal definition of the \textsc{Quadratic Unconstrained Binary Optimization} (QUBO) problem. 
Let $Q$ be an $n\times n$ matrix of weights and $x_1, x_2, \dots, x_n$ be \emph{binary} variables. Then, the QUBO problem asks for an assignment of the variables $x_i$, for $i \in [n]$ that minimizes
$$\sum_{i,j} Q_{ij}x_ix_j.$$
\begin{comment}
According to \cite{glover2018tutorial}, it is commonly assumed, that the $Q$ matrix is symmetric or in upper triangular form, which can be achieved without a loss of generality by changing $Q$ according to the algorithms:

\begin{itemize}
	\item Symmetric form: For all $i$ and $j$ except $i = j$, replace $q_{ij}$ by $(q_{ij} + q_{ji}) / 2$.
	\item Upper triangular form: For all $i$ and $j$ with $i < j$, replace $q_{ij}$ by $q_{ij} + q_{ji}$. Then replace all $q_{ij}$ for $j < i$ by 0. (If the matrix is already symmetric, this just doubles the $q_{ij}$ values above the main diagonal, and then sets all values below the main diagonal to 0).
\end{itemize}
Our implementation makes use of the upper triangular form.
\end{comment}

\subsection{Kemeny Rank Aggregation}

Let $C$ be a finite set of candidates. A \emph{vote} (or \emph{ranking}) $\pi$ over $C$ is a total order on $C$. If for two candidates $a, b \in C$, candidate $a$ is ranked better than $b$ in the vote $\pi$, then we write ${a <_\pi b}$. 
Therefore, the smallest candidate $a$ according to $<_\pi$ is the winner of the vote $\pi$ and can be interpreted as getting position $1$.
For two votes $\pi_{i}$ and $\pi_{j}$ the number of pairs of candidates which are ordered differently within these votes is called the \emph{Kendall Tau distance} of $\pi_{i}$ and $\pi_{j}$, which is written as $\KTdist(\pi_{i},\pi_{j})$. Formally, this means:

\begin{center}
	$\KTdist(\pi_{i},\pi_{j}) = |\{(c_{1},c_{2}) \in C \times C \mid {c_{1} <_{\pi_{i}} c_{2}} \land c_{1} >_{\pi_{j}} c_{2} \}|$
\end{center}

Given a ranking $\pi$ over $C$ and a set $\Pi$ of votes over $C$, the \emph{Kemeny score} of $\pi$ w.r.t.~$\Pi$, is the sum of the Kendall Tau distances between $\pi$ and each $\pi_{i} \in \Pi$. The goal of \textsc{KRA} is to compute a ranking $\pi_{\text{min}}$ of $C$, with the smallest possible Kemeny score. Such a ranking is called a \emph{Kemeny ranking}.

\section{Framework}

In this section, we aim to give a theoretical explanation of the concepts we are using. First, we present data reduction rules for \textsc{KRA}, taken from~\cite{DBLP:journals/aamas/BetzlerBN14}, that we implemented as a preprocessing step.\footnote{In \cite{DBLP:journals/aamas/BetzlerBN14} votes are interpreted in the inverse way, meaning that candidate $a$ is ranked better than $b$ in a vote if $a>b$. Hence, in the original work, the following rules are defined according to the $\geq_s$-majority.}
% in order to cut a problem instance into a collection of smaller instances.
 Then, we present and our QUBO formulation for \textsc{KRA}.

\subsection{Data Reduction Rules}
As the number of available qubits on a quantum annealer is restricted we use some known data reduction rules as a preprocessing step that cut an instance into a collection of smaller instances that can be solved independently. 
%We use two different rules taken from~\cite{DBLP:journals/aamas/BetzlerBN14} are also implemented and usable in out tool. For completeness reasons, we shortly discuss them, the original paper also contains proofs for each.

\subsubsection{$\leq_{3/4}$-Majority Rule}\label{s:redrule1}
Let $C$ be a set of candidates and $\Pi$ be as set of votes.
Further, let $a,b \in C$ be candidates. If, for all votes $\pi_{i}$ over $C$, $a <_{\pi_{i}} b$, then $a < b$ in a Kemeny ranking. Intuitively, if $a$ gets a lower ranking than $b$ in every single vote, $a$ obviously has to be placed lower than $b$ in a Kemeny ranking. 
We can generalize this idea in the following way.
Let $s \in [0, 1]$. If $a <_{\pi_{i}} b$ in at least $s\cdot |\Pi|$ many votes $\pi_i$, then we say that the pair $(a, b)$ is a \emph{clean pair} according to the $\leq_{s}$-majority and denote this by $a\leq_s b$.
If a candidate $a$ forms a clean pair according to the $\leq_{s}$-majority with every other candidate $b\in C\setminus\{a\}$, then we call $a$ a \emph{clean candidate} with respect to the $\leq_{s}$-majority.
It was shown in~\cite[Lemma~1]{DBLP:journals/aamas/BetzlerBN14} that for clean candidates with respect to the $\leq_{3/4}$-majority ($s=3/4$), every Kemeny ranking respects the clean pairs in which $a$ is involved. 
Based on this observation, a linear partial kernel with respect to the parameter $d_a$ was obtained where $d_a=\sum_{\pi_i, \pi_j\in \Pi} \KTdist(\pi_i, \pi_j) / (|C|\cdot (|C|-1))$ is the average KT-distance. The kernel is based on the following $\leq_{3/4}$-majority rule:

For a \textsc{KRA} instance $(C, \Pi$), let $N := \{n_1, \dots, n_t\}$ be the set of clean candidates with respect to the $\leq_{3/4}$-majority such that $n_i \leq_{3/4} n_{i+1}$ for $i \in [t-1]$. Then, define 
\begin{align*}
	D_0 &:= \{c \in C\setminus N \mid c \leq_{3/4} n_1\},\\
	D_i &:= \{c \in C\setminus N \mid n_i \leq_{3/4} c \wedge c \leq_{3/4} n_{i+1}\},\\
	D_t &:= \{c \in C\setminus N \mid n_t \leq_{3/4} c\},
\end{align*}
for $i \in [t-1]$. Replace the original instance by the $t+1$ sub-instances induced by $D_i$ for $0 \leq i \leq t$. 

As shown by Betzler et al., this data reduction rule is safe and preserves all Kemeny rankings, a desirable property for computing a diverse set of Kemeny rankings. Another benefit of this rule is that the individual sub-instances can be solved independently on a quantum annealer needing way fewer qubits than the original instance.

\subsubsection{Extended Condorcet Rule}\label{s:redrule2}
In the extended Condorcet rule, we move from comparing the relative position of one candidate to all other candidates to considering the relative position of a set of candidates $C'$ to the remaining candidates $C\setminus C'$. Thereby, the rule cuts an instance according to the strongly connected components of its majority graph. 
\begin{definition}[Majority graph]
	Let $(C, \Pi)$ be an instance of \textsc{KRA}. The \emph{weak (strict)} marjority graph of $(C, \Pi)$ is a directed graph with vertex set $C$ and an arc from $u$ to $v$ if and only if $u < v$ in at least (in more than) half of the votes.
\end{definition}
As our interest is in computing a diverse set of solutions, we use a version of the extended Condorcet rule which is based on the \emph{weak} majority graph and which preserves all Kemeny rankings. If the context is clear, we call this rule the Condorcet rule. It works as follows~\cite{DBLP:journals/aamas/BetzlerBN14}.
Let $(C, \Pi)$ be an election and let $C_1, C_2, \dots, C_t \subseteq C$ be the vertex sets of the strongly connected components in the weak majority graph of $(C, \Pi)$ following a topological order. Replace the original instance by the sub-instances induced by $C_i$ with $|C_i| \geq 2$, $i \in [t]$.

While this rule preserves all Kemeny rankings, it is possible that the instance obtained after applying this rule can still be reduced by the $\leq_{3/4}$-majority rule, as shown in Example~2 of~\cite{DBLP:journals/aamas/BetzlerBN14}. If we instead consider the strongly connected components of the \emph{strict} majority graph, the resulting kernel, with respect to the parameter $d_a$, cannot be further compressed by the $\leq_{3/4}$-majority rule, implying a smaller kernel size, but we might lose Kemeny rankings as pairs of candidates with exactly $1/2$ of the votes in both directions are not connected in the strict majority graph and an ordering on them must be fixed during the topological ordering. 
As our objective is to leave the computation of multiple solutions to the quantum annealer, we used the Condorcet rule based on the weak majority graph in the following.
Alternatively, we could restrict ourselves to only computing multiple solutions on the sub-instances and use classical post processing in order to compute all topological orderings of the strict majority graph and concatenate the solutions of the sub-instances accordingly. It is not hard to see that once we have the strongly connected components, computing a diverse set of topological orderings can be done by seeing the problem as an instance of the \textsc{Completion of an Ordering} problem with all costs set to 1 and using the parameterized algorithm for the diverse version proposed in~\cite{DBLP:conf/ijcai/ArrighiFLO021}.

\subsection{QUBO formulation for Kemeny Rank Aggregation}\label{s:kra-qubo}

We now discuss our QUBO formulation of \textsc{KRA}: Given a ranking over a set of $n$ candidates $C$, we use $n^2$ binary variables, denoted by $c_{i,j}$ with $1 \le i,j \le n$. Intuitively, if variable $c_{i,j} = 1$, this means candidate $i$ has position $j$. 
Binary variables might only take the values $0$ or $1$ which can be interpreted as Boolean values.
We interpret the variables $c_{i,j}$ as a two-dimensional grid. In order to obtain a valid Kemeny ranking, we have to ensure, that each candidate has exactly one position within the final ranking ($\forall i: \sum_{j = 1}^{n} c_{i,j} = 1$, we achieve this via so called row-penalties) and that each position is taken by exactly one candidate ($\forall j: \sum_{i = 1}^{n} c_{i,j} = 1$, this, in turn, is achieved via column-penalties).
A penalty is a huge positive weight that is multiplied to a term in the objective function which gets a positive value under a variable assignment that we want to forbid. 
By turning side-constraints into penalty terms, we can obtain an unconstrained QUBO instance. 
Therefore, we need to ensure that the cost of a penalty is too high to be be paid in any optimal solution.
We choose $P = \abs{C}^{2} \cdot \abs{\Pi}$ as the penalty. This choice gets clear later.
% in our implementation, which is sufficient to guarantee that penalties which are designed to keep our rules intact will always trump weights assigned to  rankings. More on that below.

\subsubsection{Row-Penalties}

The row-penalties ensure that each candidate gets assigned exactly one position. This is obtained by the following term of the final objective function. For each row $i \in [n]$ we define
\[\row_{i} = P\left(1 - (\sum_{j=1}^{n} c_{i,j})\right)^{2}\]
The minimal value for this equation is $0$, which is obtained if and only if exactly one of the $c_{i,j}$ in the sum is true. 
If for fixed $i$, none of the variables $c_{i,j}$, for $j \in [n]$ is true, then the term $\row_{i}$ evaluates to $P$. If on the other hand $k>1$ many variables $c_{i,j}$ are true, then $\row_{i}$ evaluates to $P(1-k)^2 = P(-(k-1))^2 = P(k-1)^2 \geq P$.

\subsubsection{Column-Penalties}
Similarly, the column-penalties ensure that each position $j$ is assigned with exactly one candidate. For each $j \in [n]$ we set:
%
%The column-penalties are similar to the row-penalties since they do the same thing, just for columns. For our ranking this means, that they prevent any result in which a single position is taken by anything else than exactly one candidate, so overall each candidate needs to take a position and each position needs to be taken by exactly one candidate. The self-loops are still present here in order to ensure, that it is never the best solution to not have a position be taken by any candidate at all.
\[\col_{j} = P\left(1 - (\sum_{i=1}^{n} c_{i,j})\right)^{2}\]
%\begin{figure}[H]
%	\centering
%	\scalebox{0.7}{
%	\tikz{		
%		\node (A) at (0,0) [circle,draw] {$c_{1,1}$};
%		\node (B) at (0,-2) [circle,draw] {$c_{2,1}$};
%		\node (C) at (0,-4) [circle,draw] {$c_{3,1}$};
%		
%		\draw[-] (A) to[out=30, in=330, looseness=5] (A);
%		\draw[-] (B) to[out=30, in=330, looseness=5] (B);
%		\draw[-] (C) to[out=30, in=330, looseness=5] (C);
%		\draw[-] (A) to[out=225, in=135] (B);
%		\draw[-] (A) to[out=180, in=180] (C);
%		\draw[-] (B) to[out=225, in=135] (C);
%	}
%	}
%	\caption{Graph-representation of column-penalties}
%\end{figure}
%
%As before, each vertex in such a column has to be connected to each other one.

\subsubsection{Ranking-Penalties}\label{s:rankingpenalties}

Lastly, we need to represent the actual votes themselves. This is achieved via penalties for each pair of variables $c_{i,j}, c_{i',j'}$ with $i\neq i'$ and $j \neq j'$. The goal is to give the quadratic term $c_{i,j}c_{i',j'}$ a weight that reflects how many votes are violated if both, candidate $i$ gets position $j$ and candidate $i'$ gets position $j'$. It was shown in~\cite[Sec.~5.1]{DBLP:conf/fsttcs/ArrighiFO020} that the Kemeny score of a ranking can be computed by summing over the cost of each pair of candidates, i.e., the number of votes in which the relative position of the pair is violated. This way, summing up the penalties over all pairs of candidates, we obtain the Kemeny score of the ranking associated with a variable assignment.

Let $w_{i,j}$ be the total number of votes, in which candidate $c_j$ is positioned better than candidate $c_i$. 
More formally\footnote{The bracket notation reads as follows: if $p$ is a logical proposition, then $[p]$ yields 1 if $p$ is true and else, $[p]$ yields $0$.}, $w_{i,j} = \sum_{\pi\in \Pi} [c_j <_{\pi} c_i]$. 
If we violate this relative position of candidates in an aggregated ranking, then $w_{i,j}$ is the cost we need to pay in order to place candidate $c_i$ before candidate $c_j$.
This leaves us with the following term encoding the cost of placing candidate $c_i$ before candidate $c_j$. 
For each $i, j \in [n]$ we define:
\[
\rank_{i,j} = w_{i,j} \cdot \sum_{k=1}^{n-1} \sum_{l>k}^{n} c_{i,k} c_{j,l}
\]
%Likewise, we can determine $w_{j,i}$, i.e. the total number of votes, in which candidate $j$ is positioned better than candidate $i$:
%
%\begin{equation*}
%	w_{j,i} \cdot \sum_{k=2}^{n} \sum_{l<k}^{n} c_{i,l} c_{j,k}
%\end{equation*}

%We now have 2 formulas which we can use to determine the weights of edges, however if we take the sum of those 2, it will leave us with the total amount of penalties between those two candidates in regards to relative positioning, which we will call $W_{i,j}$.
%
%\begin{equation*}
%	W_{i,j} = w_{i,j} \sum_{k=1}^{n-1} \sum_{l>k}^{n} c_{i,k} c_{j,l} + w_{j,i} \sum_{k=2}^{n} \sum_{l<k}^{n} c_{i,k} c_{j,l}
%\end{equation*}

\subsubsection{The final formulation}
%If we now also consider the row- and column-penalties, we are left with total amount of penalties overall:
We are ready to formulate the final unconstrained objective function depending on all variables $c_{i,j}$ for $i, j \in [n]$:
\[f(c_{1,1}\dots c_{n,n}) = \sum_{i=1}^n \left(\row_i + \col_i + \sum_{j=1}^n \rank_{i,j}\right)\]

Coming back to the penalty $P$ used for row- and column-penalties: Since we choose it to be $P = \abs{C}^{2} \cdot \abs{\Pi}$, we now see, that even if every single vote contradicts every single possible relative positioning 
(which is an impossible to reach upper bound) breaking the rules of each candidate taking exactly one position and each position being taken by exactly one candidate, will still come with a higher penalty, and thus will never happen in an optimal solution.

\section{Implementation}
The project is written in Python using \DWaveName's \textsc{Dimod} API.
%, since the \textsc{Dimod} API is also in Python. Dimod
%\footnote{\url{https://github.com/dwavesystems/dimod}} 
%is a shared API for samplers, it provides classes for quadratic models, such as the binary quadratic model (BQM) class that contains Ising and QUBO models used by the D-Wave samplers. 
%It also provides reference examples of samplers and composed samplers, which are useful for small, locally solvable models.

\subsection{Data-sets}\label{datatype}
We use two different data-sets in our evaluation. The first is the \emph{Formula 1} data-set from the experimental evaluation of the data reduction rules in~\cite{DBLP:journals/aamas/BetzlerBN14}.
The number of candidates in this data-set ranges from 6 to 28.
They also considered other real world data-sets such as ski rankings and web search results. But those data-sets focus on instances with up to 69, resp., 200 candidates, which are too large to be mapped onto the physical hardware of the considered quantum annealers. Hence, we exclude them from our evaluation. 

In contrast, we included a second data-set of randomly generated instances ranging from 3 to 100 candidates in order to gradually increase the instance size to find the limitations of the different solver. We created the instance by starting with a sorted list and then shuffling it with standard python methods. As expected, the randomly generated instances do not provide much structure for the data reduction rules and are used for comparing the runtime and solution quality of the different solvers with different parameters. In contrast, the Formula~1 data is used to compare the diversity of solutions and the impact of the data reduction rules.
Details on the encoding of the data, documentation of our implementation as well as instructions on how to run the experiments can be found in the supplementary material along with the code and the data-sets.

%\textcolor{red}{Initially, the real world data from \cite{DBLP:journals/aamas/BetzlerBN14} was used, but these where partial elections. Removing partial candidates and duplicates is straight forward and therefore done when importing data-sets. Additionally, a collection of elections from \href{https://www.preflib.org/data}{PrefLib} served as real world data, mainly the Netflix and the Electorial Reform Society (ERS) Election Data as well as a set of randomly generated elections.}
%
%\textcolor{blue}{Since we started with the set of input files from \cite{DBLP:journals/aamas/BetzlerBN14}, this format was initially used. It is line based and every line has to hold one vote. The vote itself is a list of candidate names, separated by the \emph{$>$} sign.
%%
%However, when we found additional input data from PrefLib, as mentioned above, in another format, we implemented support for it as well. This format is summarizing the relations of all candidate pairs over all rankings. Examples for both formats can be found with the code.}

\subsection{Preprocessing}\label{s:kernel}
%We implemented a preprocessing module that supports different data reduction modes.
We implemented the above described data reduction rules that cut instances into subinstances.
%To archive that, we require each mode to generate a list of candidate sets, such that the candidates in the $i$-th set have to have better positions in an optimal ranking as the candidates in the $j$-th set if $i < j$ holds. 
%Having this requirement enables us to generate disjoint sub-instances,
% (which we will call sub-kernels), 
%They create subinstances that can be solved independently, and whose solution can be concatenated in order to obtain a valid solution. 
If we aim for a diverse solution set, we concatenate the best solutions among the sub-instances in a cross-product manner.
% We currently have three different data reduction modes implemented: no data reduction, $\leq_{3/4}$-majority rule, and Condorcet rule. 
The available modes are: no data reduction, $\leq_{3/4}$-majority rule, and Condorcet rule. 
% The \verb|none| mode just creates one instance containing all candidates, so no data reduction is done. The \verb|candidates| mode implements the $\leq_{3/4}$-majority reduction rule and the \verb|Condorcet| mode implements the extended Condorcet reduction rule.

\begin{comment}
\subsection{Generating the QUBO formulation}
We generate a QUBO formulation for each sub-instance of size $\geq 2$ (since those of size 1 have a trivial solution) which we obtained by the procedures described above. 
\end{comment}
%If no kernelization is active, we just use the original input as one sub-instance. In order to create our QUBO formulation we use the numpy array data structure from NumPy\footnote{\url{https://numpy.org/}}. Thereby, we can generate a BQM from the Dimod library by passing the numpy array without having to create additional objects representing the same matrix. The process of generating the QUBO formulation has two steps: In the first step we set the values of the $\rank$ penalties as described above.
%%in section \ref{s:rankingpenalties}. 
%In the second step we set the values for the row and column constraints based on the $\row$ and $\col$ penalty terms.
%% formula that can be found in section \ref{s:rowcolumnformula}. 
%The code related to this process can be found in our sub-kernel class. We then use the generated BQM as an interface to different solvers discussed next.

\subsection{Solving with Different Solvers}\label{solver}

With the problem formulation finished, the BQM model can be solved by a variety of different solvers, each with different pros and cons. Generally, the solvers provided by D-Wave can be split into 3 categories: mathematical solving, simulated annealing, and real quantum annealing.

%\subsubsection{Mathematical solver}
\subsubsection{Steepest Decent}
%The \ExactSolverName solver
%%\footnote{\href{https://docs.ocean.dwavesys.com/en/stable/docs_dimod/reference/sampler_composites/samplers.html\#exact-solver}{DWave Docs - Exact Solver}} 
%was the first mathematical solver to be tested, but due to the exponential blowup, it became unusable for instances of size $>5$ because of memory consumption and computation times over 5 minutes, therefore it is omitted in the evaluation. The \ExactSolverName solver calculates the value of the objective function for every possible combination of variable assignments. Hence, it always shows a worst-case runtime behavior which is exponential in the number of variables.\todo{can be removed}

%If variable $i$ has $k_i$ cases, this results in $k_1, \dots, k_n$ cases, which grows exponentially for constant $k_i$ in the number of variables.
The first solver, \SteepestDescentSolverName, that we consider from the repertoire of D-Wave 
%The other classical local solver, the \SteepestDescentSolverName solver 
uses a steepest descent approach which is the discrete analogue of gradient descent, but the best move is computed using a local minimization rather than computing a gradient. 
%For a given input model’s graph $G=(V,E)$, $V$ being a set of graph vertices and $E$ a set of edges, runtime complexity of the underlying C++ implementation is $\mathcal{O}(|E|)$ for initialization phase and $\mathcal{O}(|V|)$ per downhill step.
For a given input QUBO on $n$ variables, the runtime complexity of the underlying C++ implementation is $\mathcal{O}(n^2)$ for the initialization phase and $\mathcal{O}(n)$ per downhill step. At the time of our experiments, \SteepestDescentSolverName did not support to compute multiple solutions.

\subsubsection{Simulated annealing}

Simulated annealing is a probabilistic technique for finding local optima of a given objective function. 
%Specifically, it is a metaheuristic to approximate global optimization in a large search space for an optimization problem, such as QUBO. 
\DWaveName provides with the \SimulatedAnnealingSamplerName solver
%\footnote{\url{https://github.com/dwavesystems/dwave-neal}} 
an implementation of a simulated annealing algorithm that approximates Boltzmann sampling.
Their approach follows the work of~\cite{kirkpatrick1983optimization}.

\subsubsection{Quantum annealing}
D-Wave provides two different types of quantum annealers: a quantum processing unit (QPU), and a hybrid model. In the hybrid model, the Ocean SDK takes care of assigning a QUBO instance to qubit bias and coupling values for the hardware graph, with the drawback of restricted access to parameters like the number of samples (\verb|num_read|) or the number of output solutions. In contrast, the QPU model allows for direct access to the quantum machine but requires the user to take care of optimal values for parameters as well as finding an embedding onto the hardware herself. The latter is a non-trivial task and without your own implementation, just a general heuristic is available that only allows to embed significantly smaller instances than possible with the hybrid model.
The most developed QPU family currently available is the Advantage model with over 5,000 qubits. In comparison with its predecessor, the 2,000Q QPU, it relies on a new topology, the \emph{Pegasus graph}, describing the connectivity of qubits on the chip. 
In our experiments we use the Advantage model. We refer to the solver based on the hybrid model as \LeapHybridSamplerName, and to the Advantage QPU as \DWaveSamplerName.

%\textcolor{red}{\DWaveName provides quantum-classical hybrid sololve arbitrary application problems formulated as quadratic models. These solvers accept arbitrarily structured, unconstrained problems formulated as BQMs, with any constraints typically represented through penalty models and run on a QPU. The superconducting QPU at the heart of the \DWaveName system, which operates at a temperature of approximately 12 $mK$, is a controllable, physical realization of the quantum Ising spin system in a transverse field such that the classical spin states represent a low energy solution. The Model \DWaveSamplerName \DWaveName 2000Q1 QPU offers 2041 Qubits and the \LeapHybridSamplerName Advantage\_system3.2 with 5556 for solving the BQMs.}

\paragraph{Local Search}
We implemented the local search algorithm described in~\cite{DBLP:conf/alenex/SchalekampZ09}. Local search was also considered in the experiments in~\cite{DBLP:journals/mss/AliM12} where it was observed to compute significantly better solutions than faster algorithms like \Borda.
The algorithms goes through the candidates in a random order. When considering position $i$, the candidate at position $i$ is moved to the position that gives the largest improvement to the current Kemeny score. The procedure stops if no element has been moved and all positions were considered.
As the algorithms uses a random order, we execute the local search algorithm multiple times in order to compute multiple solutions.

\paragraph{Borda}
We implemented the Borda algorithm from \cite{DBLP:journals/mss/AliM12}. It computes (in our notation) for each  candidate $a$ the value $q_{c_i} = \sum_{c_j \in C}w_{i,j}$ and then returns the permutation of candidates that sorts $q_{c_i}$ in descending order.
As the algorithm computes a single solution deterministically it is not able to compute a diverse set of solutions.
In~\cite{DBLP:journals/mss/AliM12} it was observed that in the case of no or strong consensus, the \Borda performs best among the considered solver. Also in the comparison~\cite{DBLP:conf/alenex/SchalekampZ09} \Borda performed very well.

\paragraph{Quick-Sort}
Several sorting based algorithms were considered in~\cite{DBLP:journals/mss/AliM12} from which the one based on quick sort performed best in our experiments.
The algorithm sorts the candidates according to the predicate $w_{ij} < w_{ji}$ that, if true, sorts candidate $c_i$ left of candidate $c_j$.
Also \QuickSort is only able to compute a single solution.

\subsection{Parameters}
We focus our analysis on the impact of two parameters. The parameter \verb|num_reads| determines the number of samples performed by the solver. Its standard value is 1. Increasing this parameter leads to better solutions but causes higher runtime. We investigate its impact in Exp.~2.
At the time of our experiments, \verb|num_reads| is supported by all solvers from \DWaveName, except \LeapHybridSamplerName.
The parameter \verb|max_answers| determines the number of solutions output by the solvers. While at the beginning of this project  \verb|max_answers| was also supported by the \LeapHybridSamplerName solver, at the time of our experiment it is only supported by \DWaveSamplerName, and \SimulatedAnnealingSamplerName.
%The solvers provided offer a variety of parameters to influence how the problem is run, e.g. how many samples are going to be collected. Some of which are highly technical, like the $annealing\_shedule$ or the $beta$. 

\begin{comment}
\subsection{Command Line Interface}\todo{can go}
We provided a configuration interface for parameters as CLI. It enables the user to specify the input data, choose the format of the input data, enable debugging, and choose the kernelization mode. Currently there are 3 kernelization modes implemented, namely \verb|none|, \verb|candidates| ($\leq_{3/4}$-majority), and \verb|condorcet|. Additionally, the user can specify a solver (from those discussed above)
%in Section~\ref{solver} 
and tweak the \verb|num_read| parameter. 
%The program can be started with the -h option to get a short overview.
\end{comment}

\section{Experimental Results}
All local solver benchmark tests were performed on a system running Linux Pop OS (Ubuntu 22.4 LTS) with a 5.17.5 kernel, an AMD Ryzen 5 2600X six core processor (12 threads) with a 3.6 GHz base clock and a 4.2 GHz boost clock speed, equipped with 32GB 3200MHz ram.
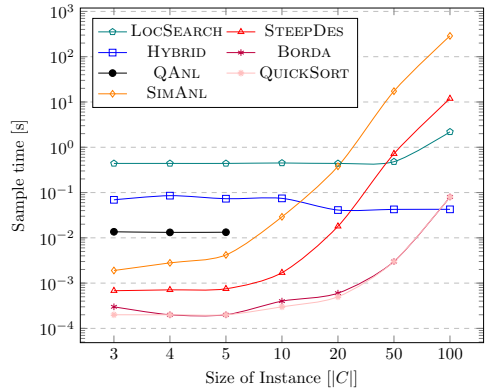
\begin{figure}[t]
	\centering
	\begin{tikzpicture}[scale=0.63]
		\centering
		\begin{axis}[
			xlabel={Size of Instance [$|C|$]},
			ylabel={Sample time [s]},
			ymode=log,
			% stupid workaround for pfgplots having large gaps in the x axis since the data doesn't grow in a linear manor 
			% 1 = 3     2 = 4   3 = 5   4 = 10  5 = 20  6 = 50  7 = 100  8 = 150     9 = 200     10 = 250 
			xtick = {1,2,3,4,5,6,7,8,9,10},
			xticklabels =  {3,4,5,10,20,50,100,150,200,250},
			legend pos=north west,
			legend columns=2,
			ymajorgrids=true,
			grid style=dashed,
			]

			%			\addplot[color=green,mark=star,smooth]
			%			coordinates {(1,0.000824)(2,0.0926)(3,113.137)};
			
			\addplot[color=teal,mark=pentagon,smooth]
			coordinates {(1,0.44)(2,0.44)(3,0.44)(4,0.45)(5,0.44)(6,0.48)(7,2.19)};
			
			\addplot[color=red,mark=triangle,smooth]
			coordinates {(1,0.00068)(2,0.00071)(3,0.00075)(4,0.0017)(5,0.018)(6,0.72)(7,11.88)};
			
			\addplot[color=blue,mark=square,smooth]
			coordinates {(1,0.0691)(2,0.0854)(3,0.0731)(4,0.0744)(5,0.0411)(6,0.0425)(7,0.0426)};    
			
			\addplot[color=purple,mark=asterisk,smooth]
			coordinates {(1,0.0003)(2,0.0002)(3,0.0002)(4,0.0004)(5,0.0006)(6,0.003)(7,0.08)};
			
			\addplot[color=black,mark=*,smooth]
			coordinates {(1,0.0136)(2,0.0132)(3,0.0133)};  
			
			\addplot[color=pink,mark=10-pointed star,smooth]
			coordinates {(1,0.0002)(2,0.0002)(3,0.0002)(4,0.0003)(5,0.0005)(6,0.003)(7,0.08)};
			
			\addplot[color=orange,mark=diamond,smooth]
			coordinates {(1,0.0019)(2,0.0028)(3,0.00419)(4,0.0291)(5,0.38)(6,17.34)(7,287.76)};

			\legend{\LocalSearch,\SteepestDescentSolverName,\LeapHybridSamplerName,\Borda,\DWaveSamplerName,\QuickSort,\SimulatedAnnealingSamplerName}
			
		\end{axis}
	\end{tikzpicture}
	\caption{Sample, respectively, QPU time with increasing number of candidates, average over 10 random instances.}
	\label{fig:exp1}
\end{figure}

\subsection{Exp.~1: Runtime Performance (Fig.~\ref{fig:exp1})}
We measure the runtime for each solver on a set of randomly generated instances of increasing size with 10 instances per size (number of candidates).
Each instance contains as many votes as candidates.
%The runtime was measured on our randomly generated instances in increasing size. 
%As already mentioned, the \ExactSolverName is omitted for larger instances due to bad performance and its ability to handle at most 18 variables. 
All solvers were initialized with default parameters, and only the pure sample time is measured. For the \LeapHybridSamplerName solver, the QPU time was measured. 

%\resizebox{225pt}{200pt}{
%}

We observe that the sampling time of \DWaveSamplerName, \LeapHybridSamplerName, and \LocalSearch stays rather constant, while the runtime for \SimulatedAnnealingSamplerName, \SteepestDescentSolverName, \Borda, and \QuickSort first outperforms the other solver but then increases significantly with larger instance sizes.
%For instances with size $|C| \le 20$ \SteepestDescentSolverName is outperforming all other solvers. \SimulatedAnnealingSamplerName has a slightly higher sampling time. The \LeapHybridSamplerName solver is faster than any other solver for instances with size $|C| > 20$. \todo{eval new heuristics}
%This is omitting all overhead, especially network traffic and waiting in the queues for the QPU access. 

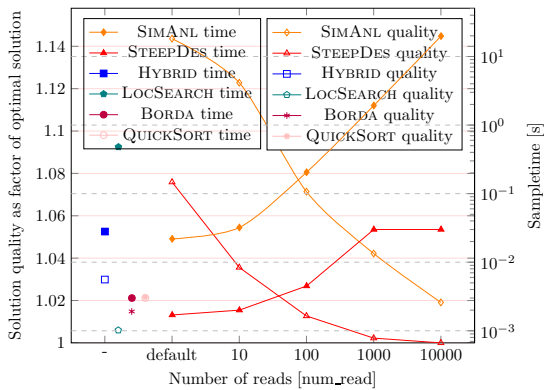
\begin{figure}
	\centering
\begin{tikzpicture}[scale=0.63]
	\begin{axis}[
		xlabel={Number of reads [num\_read]},
		ylabel={Solution quality as factor of optimal solution},
		% stupid workaround for pfgplots having large gaps in the x axis since the data doesn't grow in a linear manor 
		% 1 = 3     2 = 4   3 = 5   4 = 10  5 = 20  6 = 50  7 = 100  8 = 150     9 = 200     10 = 250 
		xtick = {0,1,2,3,4,5},
		xticklabels =  {-,default,10,100,1000,10000},
		legend pos=north east,
		ymajorgrids=true,
		grid style={red!15},
		ymin=1,
		]
		\addplot[color=orange,mark=diamond,smooth]
		coordinates {(1,1.1436)(2,1.1228)(3,1.0714)(4,1.0422)(5,1.0191)};
		
		\addplot[color=red, mark=triangle,]
		coordinates {(1,1.0759)(2,1.03561)(3,1.0127)(4,1.00225)(5,1)};
		
		\addplot[color=blue,mark=square,]
		coordinates {(0,1.0299)};
		
		\addplot[color=teal,mark=pentagon,]
		coordinates {(0.2,1.005985)};
		
		\addplot[color=purple,mark=asterisk,]
		coordinates {(0.4,1.01483)};
		
		\addplot[color=pink,mark=10-pointed star,]
		coordinates {(0.6,1.02155)};
		
		\legend{\SimulatedAnnealingSamplerName quality, \SteepestDescentSolverName quality,\LeapHybridSamplerName quality,
		\LocalSearch quality,
		\Borda quality,
		\QuickSort quality}
		
	\end{axis}
	\begin{axis}[
		axis y line*=right,
		axis x line=none,
		ylabel={Sampletime [s]},
		ymode=log,
		% stupid workaround for pfgplots having large gaps in the x axis since the data doesn't grow in a linear manor 
		% 1 = 3     2 = 4   3 = 5   4 = 10  5 = 20  6 = 50  7 = 100  8 = 150     9 = 200     10 = 250 
		xtick = {0,1,2,3,4},
		xticklabels =  {-,10,100,1000,10000},
		legend pos=north west, legend style={fill=none},
		ymajorgrids=true,
		grid style=dashed,
		]	
		\addplot[color=orange,mark=diamond*,smooth]
		coordinates {(1,0.0219)(2,0.0321)(3,0.2054)(4,1.927)(5,19.83)};
		
		\addplot[color=red,mark=triangle*,]
		coordinates {(1,0.0017)(2,0.002)(3,0.0045)(4,0.03)(5,0.03)};
		
		% \addplot[color=green,mark=+,] coordinates {(-1,23.9)};
		
		\addplot[color=blue,mark=square*,]
		coordinates {(0,0.028)};
		
		\addplot[color=teal,mark=pentagon*,]
		coordinates {(0.2,0.48)};
		
		\addplot[color=purple,mark=otimes*,]
		coordinates {(0.4,0.003)};
		
		\addplot[color=pink,mark=o,]
		coordinates {(0.6,0.003)};

			\legend{\SimulatedAnnealingSamplerName time,\SteepestDescentSolverName time,\LeapHybridSamplerName time,
			\LocalSearch time,
			\Borda time, 
			\QuickSort time}

	\end{axis}
\end{tikzpicture}
\caption{Solution quality as a factor of the optimal solution, average over 10 random instances with 20 candidates.}
\label{fig:exp2}
\end{figure}

\subsection{Exp.~2: Solution Quality (Fig.~\ref{fig:exp2})}
Since a default sample size of 1 is used for \DWaveSamplerName, \SimulatedAnnealingSamplerName, and \SteepestDescentSolverName, larger instances are not going to be solved optimally. In fact, solution quality is greatly decreasing with increasing instance size. To improve solution quality, with the parameter \verb|num_read| we can set the number of samples taken. In order to measure solution quality, the cost of an optimal solution is calculated by an Integer Linear Program and compared with the Kemeny score of the best solution found by the different solvers. 
We measured average solution quality and average runtime for all random instances of size 20, with increasing \verb|num_read|.
For comparison, also the other solvers is visualized, although they do not support the parameter. The \DWaveSamplerName solver was excluded from this experiment as it could not solve the instances of size 20 and only showed trivial behavior on smaller sized instances.

%\resizebox{225pt}{200pt}{
%}

\SimulatedAnnealingSamplerName and \SteepestDescentSolverName show a clear trade-off between solution quality and sample time. With the default parameters, \LeapHybridSamplerName outperforms \SimulatedAnnealingSamplerName and \SteepestDescentSolverName in solution quality. With increasing \verb|num_read|, \SimulatedAnnealingSamplerName and \SteepestDescentSolverName do outperform \LeapHybridSamplerName in terms of solution quality. For \SimulatedAnnealingSamplerName this only happens with a huge increase in sample time. With $\verb|num_read|=100$, \SteepestDescentSolverName outperforms \LeapHybridSamplerName in solution quality and sample time. Increasing \verb|num_read| lets \SteepestDescentSolverName find the optimal solution with a slightly higher sample time than \LeapHybridSamplerName. 
From those solvers allowing to compute multiple solutions, \LocalSearch shows the best solution quality under the standard parameters but also has the highest runtime.
%The exact solver provides optimal solutions at the cost of exponential time and space complexity and is therefore omitted. 
%The \SimulatedAnnealingSamplerName solver offers a clear trade-off between runtime and solution quality. With only 10 reads, the solutions offered have a 12.3\% higher Kemeny score on average while taking around $1$s. Increasing the number of samples to 100 only takes around $4.27$s while improving quality to only 7.14\% over the optimal cost. At 10,000 reads, the solution quality improves to 1.9\% over optimal cost but requires around $5$min computation time. Larger \verb|num_read| would further increase average solution quality but in contrast, the \SteepestDescentSolverName solver is faster at 10,000 reads with $0.03s$ on average and finding the optimal solution.
%The \LeapHybridSamplerName solver is only slightly worse in terms of solution quality at 4.5\% over optimal with a QPU access time of only $0.014$s.
%The total runtime, including network and queue times for the \LeapHybridSamplerName is $23.9s$ but the QPU access time is only $0.14s$, so the compute time required to solve the instance is comparable to this.

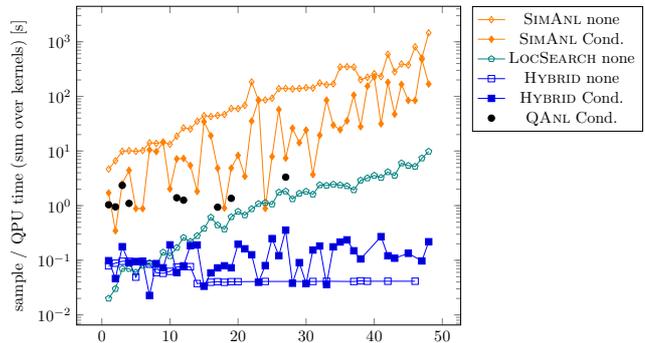
\begin{figure}[tb]
	\centering
	\begin{tikzpicture}[scale=0.6]
		\begin{semilogyaxis}[legend pos=outer north east,ylabel={sample / QPU time (sum over kernels) [s]}]
			\addplot+ [
			mark=diamond, color=orange,
			error bars/.cd,
			x dir=both, x explicit,
			y dir=both, y explicit,
			] table [x=Index, y=SumTimeKer
			] {data/SimulatedAnnealingNone.csv};
			\addlegendentry{\SimulatedAnnealingSamplerName none}
			\addplot+ [
			mark=diamond*, color=orange, every mark/.append style={solid, fill=orange},
			error bars/.cd,
			x dir=both, x explicit,
			y dir=both, y explicit,
			] table [x=Index, y=SumTimeKer
			] {data/SimulatedAnnealingCondorcet.csv};
			\addlegendentry{\SimulatedAnnealingSamplerName Cond.}
			\addplot+ [
			mark=pentagon, color=teal,
			error bars/.cd,
			x dir=both, x explicit,
			y dir=both, y explicit,
			] table [x=Index, y=Time
			] {data/LocalSearchNone.csv};
			\addlegendentry{\LocalSearch none}
			\addplot+ [
			mark=square, color=blue,
			error bars/.cd,
			x dir=both, x explicit,
			y dir=both, y explicit,
			] table [x=Index, y=SumTimeQPU
			] {data/HybridNoneClean.csv};
			\addlegendentry{\LeapHybridSamplerName none}
			\addplot+ [
			mark=square*, color=blue, 
			error bars/.cd,
			x dir=both, x explicit,
			y dir=both, y explicit,
			] table [x=Index, y=SumTimeQPU
			] {data/HybridCondorcetClean.csv};
			\addlegendentry{\LeapHybridSamplerName Cond.}
			\addplot+ [
			only marks, mark=*, color=black, every mark/.append style={solid, fill=black},
			error bars/.cd,
			x dir=both, x explicit,
			y dir=both, y explicit,
			] table [x=Index, y=SumTimeQPU
			] {data/DWaveCondorcetClean.csv};
			\addlegendentry{\DWaveSamplerName Cond.}
		\end{semilogyaxis}
	\end{tikzpicture}
	\caption{Sample time, respectively, QPU time with and without Condorcet rule.}
	\label{fig:exp3-3}
\end{figure}

\subsection{Exp.~3: Diversity and Preprocessing (Fig.~\ref{fig:exp3-1} - \ref{fig:exp3-3})}
Next, we measure the solution quality, diversity of solutions, and runtime for those solver that can compute a set of solution at once. More precisely, we compare the annealing based solvers \SimulatedAnnealingSamplerName, \LeapHybridSamplerName, \DWaveSamplerName, and \LocalSearch with and without preprocessing. Recall that \Borda, \QuickSort, and \SteepestDescentSolverName do not support to output multiple solutions on a single run of the solver and are hence not included in this experiment. 
As our previously considered randomized data do not contain enough structure for the data reduction rules, we use the Formula~1 data.
%, all other data sets from there were too big to be solvable on the quantum annealers.
We observe that on the Formula~1 data, the $\leq_{3/4}$-majority rule never applies. Hence, we excluded it from the evaluation presented here.
For the \SimulatedAnnealingSamplerName and \DWaveName all experiments were performed with $\verb|num_read|=10,000$ and up to the best 10 solutions were taken into the solution set on which we measure the diversity as the pairwise minimal $\KTdist$ and the average $\KTdist$.
For \LocalSearch, we run the solver 50 times and take the 10 best solutions found into account. 
The depicted runtime is the total runtime of all 50 runs.
Not all instances were solvable on the quantum annealers. When a data point is missing, this instance was either not embeddable onto the hardware or an invalid solution, breaking a row or column constraint, was returned. The instances are sorted by the number of candidates in increasing order. As in~\cite{DBLP:journals/aamas/BetzlerBN14} we removed invalid votes containing duplicated candidates and candidates not appearing in every vote.
In contrast to \SimulatedAnnealingSamplerName, \LocalSearch did not always found 10 different solutions. If D-Wave could solve the instance at all, it nearly always output 10 different solutions.
\begin{figure}[tb]
	\centering
	\begin{tikzpicture}[scale=0.63]
		\begin{axis}[legend pos=north west,ylabel={Percentage over optimal solution [\%]},legend style={fill=none}]
			\addplot+ [
			mark=square, color=blue,
			error bars/.cd,
			x dir=both, x explicit,
			y dir=both, y explicit,
			] table [x=Index, y=SolDeltaPercent
			] {data/HybridNone.csv};
			\addlegendentry{\LeapHybridSamplerName none}
			\addplot+ [
			mark=square*, color=blue, every mark/.append style={solid, fill=blue},
			error bars/.cd,
			x dir=both, x explicit,
			y dir=both, y explicit,
			] table [x=Index, y=SolDeltaPercent
			] {data/HybridCondorcet.csv};
			\addlegendentry{\LeapHybridSamplerName Condorcet}
			\addplot+ [
			mark=diamond, color=orange,
			error bars/.cd,
			x dir=both, x explicit,
			y dir=both, y explicit,
			] table [x=Index, y=SolDeltaPercent
			] {data/SimulatedAnnealingNone.csv};
			\addlegendentry{\SimulatedAnnealingSamplerName none}
			\addplot+ [
			mark=diamond*, color=orange,
			error bars/.cd,
			x dir=both, x explicit,
			y dir=both, y explicit,
			] table [x=Index, y=SolDeltaPercent
			] {data/SimulatedAnnealingCondorcet.csv};
			\addlegendentry{\SimulatedAnnealingSamplerName Condorcet}
			\addplot+ [
			mark=pentagon, color=teal,
			error bars/.cd,
			x dir=both, x explicit,
			y dir=both, y explicit,
			] table [x=Index, y=SolDelta
			] {data/LocalSearchNone.csv};
			\addlegendentry{\LocalSearch none}
			\addplot+ [
			mark=*, color=black, every mark/.append style={solid, fill=black},
			error bars/.cd,
			x dir=both, x explicit,
			y dir=both, y explicit,
			] table [x=Index, y=SolDeltaPercent
			] {data/DWaveCondorcetClean.csv};
			\addlegendentry{\DWaveSamplerName Condorcet}
		\end{axis}
	\end{tikzpicture}
	\caption{Depicted is by how much \% the Kemeny score of the best solution found is above the optimal solution.}
	\label{fig:exp3-1}
\end{figure}
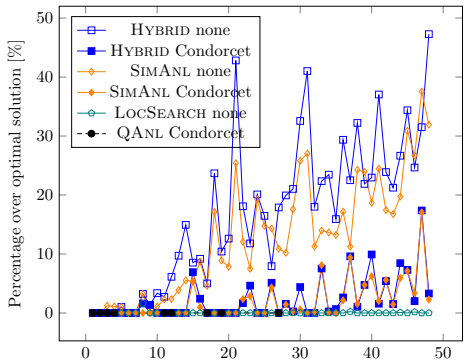
\begin{figure}[tb]
	\centering
	\begin{tikzpicture}[scale=0.63]
		\begin{axis}[legend pos=north west, ylabel={$\KTdist$ among diverse solution}] 
			\addplot+ [
			mark=diamond, color=orange,
			error bars/.cd,
			x dir=both, x explicit,
			y dir=both, y explicit,
			] table [x=Index, y=MinimalKTDistance
			] {data/SimulatedAnnealingNone.csv};
			\addlegendentry{\SimulatedAnnealingSamplerName none min KT}
			\addplot+ [
			mark=diamond*, color=orange, every mark/.append style={solid, fill=orange},
			error bars/.cd,
			x dir=both, x explicit,
			y dir=both, y explicit,
			] table [x=Index, y=MinimalKTDistance
			] {data/SimulatedAnnealingCondorcet.csv};
			\addlegendentry{\SimulatedAnnealingSamplerName Condorcet min KT}
			\addplot+ [
			mark=diamond, color=blue,
			error bars/.cd,
			x dir=both, x explicit,
			y dir=both, y explicit,
			] table [x=Index, y=AvgKTDistance
			] {data/SimulatedAnnealingNone.csv};
			\addlegendentry{\SimulatedAnnealingSamplerName none avg.~KT}
			\addplot+ [
			mark=diamond*, color=blue,
			error bars/.cd,
			x dir=both, x explicit,
			y dir=both, y explicit,
			] table [x=Index, y=AvgKTDistance
			] {data/SimulatedAnnealingCondorcet.csv};
			\addlegendentry{\SimulatedAnnealingSamplerName Condorcet avg.~KT}	
			\addplot+ [
			mark=pentagon, color=teal,
			error bars/.cd,
			x dir=both, x explicit,
			y dir=both, y explicit,
			] table [x=Index, y=AvgKTDistance
			] {data/LocalSearchNone.csv};
			\addlegendentry{\LocalSearch none avg.~KT}
			\addplot+ [
			mark=*, color=black, every mark/.append style={solid, fill=orange},
			only marks,
			error bars/.cd,
			x dir=both, x explicit,
			y dir=both, y explicit,
			] table [x=Index, y=MinimalKTDistance
			] {data/DWaveCondorcetClean.csv};
			\addlegendentry{\DWaveSamplerName Condorcet min KT}
			\addplot+ [
			mark=*, color=black, every mark/.append style={solid, fill=blue},
			only marks,
			error bars/.cd,
			x dir=both, x explicit,
			y dir=both, y explicit,
			] table [x=Index, y=AvgKTDistance
			] {data/DWaveCondorcetClean.csv};
			\addlegendentry{\DWaveSamplerName Condorcet avg.~KT}
		\end{axis}
	\end{tikzpicture}
	\caption{Minimal and average pairwise $\KTdist$ over the best 10 solutions found with and without Condorcet rule.}
	\label{fig:exp3-2}
\end{figure}
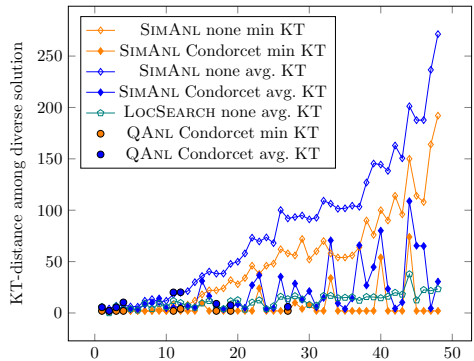

\paragraph{Observations}
We observe that the $\leq_{3/4}$-majority rule does not reduce any instance. In contrast, the Condorcet rule cuts $1/3$ of the instances into two sub-instances and $1/6$ of the instances into 3 sub-instances. Further, it reduces the size of nearly every instance. While the \LeapHybridSamplerName solver was able to solve nearly every instance in both preprocessing modes, the \DWaveSamplerName could only solve some instances after applying the Condorcet rule.
Applying the Condorcet rule increases the solution quality for all solvers significantly. With this rule, most of the instances were solved optimally by all solvers. Surprisingly with respect to solution quality, without preprocessing, the simulated and non-simulated annealers performed similarly. Among the considered 10 best solutions, the Kemeny score of the solutions differed only slightly. This means that similarly good different solutions were found. With respect to the diversity, applying the Condorcet rule reduced the minimal and average $\KTdist$ for \SimulatedAnnealingSamplerName significantly. In contrast, for the \DWaveSamplerName solver, the average $\KTdist$ seems to not be affected by the Condorcet rule. One could observe that the Condorcet rule allows for more instances to be solved with the \DWaveSamplerName with a better solution quality and a higher diversity, but those observations need to be considered with caution as we were only able to solve a few instances on the plain quantum annealer.
With respect to sample time, the QPU time of \LeapHybridSamplerName was significantly shorter than the sample time of \SimulatedAnnealingSamplerName. While the Condorcet rule reduces the total sampling time for \SimulatedAnnealingSamplerName, it increases the total QPU time for \LeapHybridSamplerName. We explain this by the increased charging time for the hardware couples which is the biggest part of the QPU time. As multiple sub-instances need to be solved, the hardware must be charged multiple times. We explain the higher QPU time of \DWaveSamplerName in contrast to \LeapHybridSamplerName with the high number of reads for \DWaveSamplerName in contrast to the unaccessible standard value for \LeapHybridSamplerName. When setting \verb|num_read|= 100, \DWaveSamplerName showed similar QPU times as \LeapHybridSamplerName with only a slight decrease in solution quality.

Even without preprocessing, \LocalSearch nearly always found the optimal solution, but only in 60\% of the instances it found 10 different solutions, while \SimulatedAnnealingSamplerName did so in 90\% of the instances. The Kemeny score of the 10th best solutions differed only slightly between \SimulatedAnnealingSamplerName and \LocalSearch. In terms of runtime, \LocalSearch was one magnitude faster than \SimulatedAnnealingSamplerName while \LeapHybridSamplerName was the fastest on nearly all instances showing a constant runtime. If we applied the Condorcet rule as a preprocessing step to \LeapHybridSamplerName, it found in 50\% of the instances with more than 15 candidates similar good solutions as \LocalSearch.

\section{Conclusion}
The property of quickly performing a huge number of samples of the solution space can become one of the main advantages of Quantum Annealing. Our experiments indicate that with Quantum Annealing, we can compute a \emph{set of solutions} that shows quite good diversity with a very good runtime behavior. But the full potential of Quantum Annealing can only be examined once the hardware evolved further and allows to solve larger instances. Until then, combining Quantum Annealing with data reduction rules as a preprocessing step can lead to promising near-term applications.

\bibliographystyle{apalike}
\bibliography{bib-short}

\end{document}